\documentclass[12pt]{article}
\usepackage{amsmath,amssymb,bm,epsfig}

\renewcommand{\thefootnote}{\fnsymbol{footnote}}
\def\slash#1{\not\!\!#1}

\begin{document}

\title{
\begin{flushright}
\begin{minipage}{0.2\linewidth}
\normalsize 
WU-HEP-16-08 \\
EPHOU-16-005
 \\*[50pt]
\end{minipage}
\end{flushright}
{\Large \bf 
{Flavor structure in $SO(32)$ heterotic string theory
\\*[20pt]}}}

\author{Hiroyuki~Abe$^{1,}$\footnote{
E-mail address: abe@waseda.jp},\ \ 
Tatsuo~Kobayashi$^{2}$\footnote{
E-mail address:  kobayashi@particle.sci.hokudai.ac.jp}, \ \ 
Hajime~Otsuka$^{1,}$\footnote{
E-mail address: h.otsuka@aoni.waseda.jp},  \\
Yasufumi~Takano$^{2}$\footnote{
E-mail address: takano@particle.sci.hokudai.ac.jp}, \ and \
Takuya~H.~Tatsuishi$^{2}$\footnote{
E-mail address: t-h-tatsuishi@particle.sci.hokudai.ac.jp}
\\*[20pt]
$^1${\it \normalsize 
Department of Physics, Waseda University, 
Tokyo 169-8555, Japan} \\
$^2${\it \normalsize 
Department of Physics, Hokkaido University, Sapporo 060-0810, Japan} \\*[50pt]}

\date{
\centerline{\small \bf Abstract}
\begin{minipage}{0.9\linewidth}
\medskip 
\medskip 
\small
We study the flavor structure in $SO(32)$ heterotic string theory on 
six-dimensional torus with magnetic fluxes.
In particular, we focus on models with the flavor symmetries $SU(3)_f$ and $\Delta(27)$. 
In both models, we can realize the realistic quark masses and mixing angles.
\end{minipage}
}

\begin{titlepage}
\maketitle
\thispagestyle{empty}
\clearpage
\tableofcontents
\thispagestyle{empty}
\end{titlepage}

\renewcommand{\thefootnote}{\arabic{footnote}}
\setcounter{footnote}{0}
\vspace{35pt}

\section{Introduction}

Superstring theory is the promising candidate for 
unified theory to describe  all the interactions including gravity 
and matter such as quarks and leptons, and Higgs fields.
Superstring theory predicts six-dimensional (6D) compact space 
in addition to the four-dimensional (4D) spacetime, i.e., 
totally the ten-dimensional (10D) spacetime.  
Massless spectrum is completely determined at the 
perturbative level when one fixes concretely a compactification, 
i.e., geometrical and gauge background.
Actually, various interesting models have been constructed 
and those include the gauge symmetry of the stanard model (SM),
$SU(3)_C \times SU(2)_L \times U(1)_Y$ and three chiral generations 
of quarks and leptons.
(See for review \cite{Ibanez:2012zz}.)
In some models, supersymmetry (SUSY) remains in 4D, while 
SUSY is broken in other models.
Thus, there are lots of (semi-)realistic models from 
the viewpoint of massless spectra.
The next issue to study about these models is to examine whether those models 
can lead numerically realistic results on 
the parameters in the SM, e.g. experimental values of 
gauge couplings and Yukawa couplings, the Higgs potential, 
the CP phase, etc.

Recently, $SO(32)$ heterotic string theory on toroidal compactification 
with magnetic fluxes was studied.
Several models with the SM gauge group and 
three chiral generations have been constructed \cite{Abe:2015mua}.
In addition, one of interesting aspects in this type of models is that 
they lead to non-universal gauge couplings among the $SU(3)_C$, $SU(2)_L$ and $U(1)_Y$ 
groups and such non-universal corrections depend on 
magnetic fluxes and K\"ahler moduli \cite{Blumenhagen:2005ga}.
Then, it is possible that these models with the SM gauge group and 
three chiral generations lead to the gauge couplings consistent with 
the experimental values \cite{Abe:2015xua}.
Note that the $E_8 \times E_8$ heterotic string theory on 
toroidal compacfitication  can not lead 
to such non-universal gauge couplings between $SU(3)_C$ and $SU(2)_L$ only by magnetic fluxes.\footnote{
See for 10D super $E_8$ Yang-Mills model on torus and orbifold with magnetic fluxes, 
e.g. \cite{Choi:2009pv}.}
Hence, this non-universality is an interesting aspect in $SO(32)$ heterotic string theory, 
although one-loop threshold corrections can lead to non-universal effects on gauge couplings 
in $E_8 \times E_8$ hetetrotic string theory \cite{Kaplunovsky:1987rp,Dixon:1990pc,Derendinger:1991hq}.
(See for numerical studies \cite{Ibanez:1991zv,Kawabe:1994mj}.)

As the next step, here we study quark and lepton masses and mixing angles 
in $SO(32)$ heterotic string theory on toroidal compactification 
with magnetic fluxes.
Because of magnetic fluxes, zero-mode profiles are non-trivially quasi-localized.
When zero-modes are localized close to each other, their couplings are 
strong.
On the other hand, when they are localized far away from each other, 
their couplings are suppressed.
Indeed, their couplings are given by the Jacobi $\vartheta$ function \cite{Cremades:2004wa}.
Thus, we could lead to phenomenological interesting results 
on fermion mass matrices.\footnote{
See for a similar studies on magnetized brane models \cite{Abe:2008sx}.}
Already, the flavor structure of the $SO(32)$ heterotic string theory on magnetized torus 
has been studied in \cite{Abe:2015mua}, 
and it is shown that several flavor symmetries appear such as 
$SU(3)_f$, $\Delta(27)$, etc.
Appearance of discrete flavor symmetries such as $\Delta(27)$, $\Delta(64)$ and $D_4$ 
have been pointed out in heterotic orbifold models \cite{Kobayashi:2004ya,Kobayashi:2006wq} and 
intersecting/magnetized D-brane models \cite{Abe:2009vi,BerasaluceGonzalez:2012vb}, and 
certain non-Abelian flavor smmetries are interesting to realize fermion masses 
and mixing angles \cite{Altarelli:2010gt,Ishimori:2010au,King:2013eh}.
Thus, we study quark masses and mixing angles, which are 
derived from $SO(32)$ heterotic string theory on toroidal compactification 
with magnetic fluxes.
We focus on models with the flavor symmetries $SU(3)_f$ and $\Delta(27)$.
We also discuss the lepton sector.

This paper is organized as follows.
In section 2, we review 
$SO(32)$ heterotic string theory on toroidal compactification 
with magnetic fluxes, and explain models with the flavor symmetries $SU(3)_f$ and $\Delta(27)$.
In section 3, we study quark masses and mixing angles 
in $SU(3)_f$ and $\Delta(27)$ models.
In section 4, we also discuss  the lepton sector 
and neutrino and Higgs masses.
Section 5 is conclusion and discussion.

\section{10D SO(32) SYM on magnetized tori}
In this section, we give a brief review of $SO(32)$ heterotic string theory on 
the torus compactification  with background magnetic fluxes.
We also explain their flavor symmetries and Yukawa couplings.

\subsection{Three generation models from $SO(32)$ heterotic string theory}

The low-energy effective field theory of $SO(32)$ heterotic string theory 
is described by 10D $SO(32)$ super Yang-Mills (SYM) theory coupled with supergravity.
We compactify the 6D space to three 2-tori $(T^2)_1 \times (T^2)_2 \times (T^2)_3$ 
with magnetic fluxes.

 We break  $SO(32)$ gauge group by inserting $U(1)$ magnetic fluxes,
\begin{equation}
	SO(32) \rightarrow SU(3)_C \times SU(2)_L \times \Pi_{a=1}^{13} U(1)_a.
\end{equation}
Since $SO(32)$ has 16 Cartan elements $H_i \; (i=1,\ldots,16)$,
we define Cartan elements of $SU(3)$ along $H_1-H_2,H_1+H_2-2H_3$ and $SU(2)$ as $H_5-H_6$.
We set Cartan elements of $U(1)_a$ as
\begin{equation}
	\begin{array}{cl}
		U(1)_1: & \frac{1}{\sqrt{2}} (0,0,0,0,1,1;0,0,\ldots,0), \\
		U(1)_2: & \frac{1}{2} (1,1,1,1,0,0;0,0,\ldots,0), \\
		U(1)_3: & \frac{1}{\sqrt{12}}(1,1,1,-3,0,0;0,0,\ldots,0), \\
		U(1)_4: & (0,0,0,0,0,0;1,0,\ldots,0), \\
		U(1)_5: & (0,0,0,0,0,0;0,1,\ldots,0), \\
		\vdots \\
		U(1)_{13}: & (0,0,0,0,0,0;0,0,\ldots,1), \\
	\end{array}
\end{equation}
in the basis $H_i$. Then, we use the basis that non-zero roots have charges
\begin{equation}
	(\underline{\pm 1,\pm 1,0,\ldots,0}),
\end{equation}
under $H_i \; (i=1,\ldots,16)$, where the underline means any possible permutations.
The gauge group enhances to a larger one if $U(1)$ fluxes are absent or degenerate.
For example, if magnetic flux along $U(1)_3$ is absent,
$SU(3)_C$ and $U(1)_3$ enhance to $SU(4)$ with Cartan elements along
$H_1-H_2,H_1+H_2-2H_3,H_1+H_2+H_3-3H_4$, in our model building.
Those enhanced symmetries can be broken by Wilson lines.

We define three 2-tori $(T^2)_i \simeq \bm{C}/ \Lambda_i$ with $i = 1,2,3$, where $\Lambda_i$ are two dimensional lattices generated by $e_1=2\pi R_i$ and $e_2=2 \pi R_i \tau_i$, $\tau_i \in \bm{C}$.
 $R_i$ and $\tau_i$ are the radii and complex structure moduli.
Then, the 6D metric  is given by
\begin{equation}
	d s_6^2=g_{mn} dx^m dx^n = 2h_{i \bar{j}} dz^i dz^{\bar{j}}, \nonumber
\end{equation}
\begin{equation}
	\begin{array}{ll}
		 g_{mn}= 
		 \begin{pmatrix}
		 	g^{(1)} & 0 & 0 \\
		 	0 & g^{(2)} & 0 \\
		 	0 & 0 & g^{(3)}
		 \end{pmatrix},
		 &
		 h_{i \bar{j}}=
		 \begin{pmatrix}
		 	h^{(1)} & 0 & 0 \\
		 	0 & h^{(2)} & 0 \\
		 	0 & 0 & h^{(3)}
		 \end{pmatrix},
	\end{array}
\end{equation}
where 
\begin{equation}
	\begin{array}{ll}
		g^{(i)} = (2 \pi R_i)^2
		\begin{pmatrix}
			1 & {\rm Re} \tau_i \\
			{\rm Re} \tau_i & |\tau_i|^2
		\end{pmatrix},
		&
		h^{(i)} = (2 \pi R_i)^2
		\begin{pmatrix}
			0 & 1/2 \\
			1/2 & 0
		\end{pmatrix}
	\end{array}
\label{eq:metrics},
\end{equation}
with the real coordidates $x^m$ for $(m,n=4,\ldots,9)$ 
and the complex coordistes $z^i=x^{2+2i}+\tau^i x^{3+2i}$ $(i=1,2,3)$ of the 6D space.
We expand $U(1)_a$ magnetic fluxes in the compact space $\bar{f}_a$ with $a=1,\ldots,13$ in the basis of K\"{a}hler forms, $w_i = idz^i \wedge d \bar{z}^i/(2 {\rm Im} \tau_i)$,
\begin{equation}
	\bar{f}_a=2 \pi d_a \sum_{i=1}^3 m_a^i w_i,
\label{eq:magnetic fluxes1}
\end{equation}
where $d_a$ are normalization factors, $m_a^i$ are integers or half-integers determined by Dirac quantization condition.

The 10D gauge fields and gaugino fields are decomposed as
\begin{equation}
	\begin{array}{ccl}
		\lambda (x^\mu,z^i) & = & \displaystyle\sum_{\ell,m,n} \chi_{\ell mn} (x^\mu) \otimes \psi^{1}_\ell (z^1) \otimes \psi^{2}_m (z^2) \otimes \psi^{3}_n (z^3), \\
		A_M (x^\mu,z^i) & = & \displaystyle\sum_{\ell,m,n} \varphi_{\ell mn,M} (x^\mu) \otimes \phi^{1}_{\ell,M} (z^1) \otimes \phi^{2}_{m,M} (z^2) \otimes \phi^{3}_{n,M} (z^3),
	\end{array}
\end{equation}
where $M=0,1,\cdots,9$ , $\mu = 0,1,2,3$ and  $\phi^i_{\ell,M} (z^i)$  and $\psi^i_\ell (z^i)$
corresponds to the $\ell$-th mode on the $i$-th $T^2$.
The $\psi^i_{\ell} (z^i)$  is the 2D spinor, and we 
denote zero-mode $\psi^i_{0} (z^i)$ as
\begin{equation}
	\psi_0^i (z^i)=
	\begin{pmatrix}
		\psi_+^i (z^i) \\
		\psi_-^i (z^i)
	\end{pmatrix}.
\end{equation}

Magnetic fluxes (\ref{eq:magnetic fluxes1}) can be obtained from the $U(1)_a$ vector potentials
\begin{equation}
	A_a^i (z^i)= \frac{\pi m_a^i}{{\rm Im} \tau_i} {\rm Im} ((\bar{z}^i+\bar{\zeta}_a^i) dz^i).
\end{equation}
Note that we included the degree of freedom of complex Wilson lines $\zeta_a^i = \zeta_a^{x^{2+2i}}+ \tau_i \zeta_a^{x^{3+2i}}$.

We use the following Gamma matrices on $(T^2)_i$,
\begin{equation}
	\begin{array}{ll}
		\Gamma^1_i =
		\begin{pmatrix}
			0 & 1 \\
			1 & 0
		\end{pmatrix},
		&
		\Gamma^2_i =
		\begin{pmatrix}
			0 & -i \\
			i & 0
		\end{pmatrix}
	\end{array},
\end{equation}
satisfying the Clifford algebra, $\left\{ \Gamma^a_i , \Gamma^b_i \right\} =2 \delta^{ab}$.
In holomorphic coordinates, then, we obtain
\begin{equation}
	\begin{array}{ll}
		\Gamma^{z^i} = (2 \pi R^i)^{-1}
		\begin{pmatrix}
			0 & 2 \\
			0 & 0
		\end{pmatrix},
		&
		\Gamma^{z^i} = (2 \pi R^i)^{-1}
		\begin{pmatrix}
			0 & 0 \\
			2 & 0
		\end{pmatrix}
	\end{array},
\end{equation}
from Eq.~(\ref{eq:metrics}).

The Dirac equation for the zero-modes with the representation $A$ and the $U(1)_a$ charge $q_a^A$ is given by
\begin{equation}
	i \slash{D}_i \psi_0^i (z^i) = i (\Gamma^{z^i} \nabla_{z_i}+\Gamma^{\bar{z}^i} \nabla_{\bar{z}_i} ) \psi_0^i (z^i)=0,
\end{equation}
with the covariant derivatives
\begin{equation}
	\begin{array}{l}
		\nabla_{z^i} = \partial_{z^i}-i q_a^A (A_a^i)_{z^i}, \\
		\nabla_{\bar{z}^i} = \partial_{\bar{z}^i}-iq_a^A (A_a^i)_{\bar{z}^i}.
	\end{array}
\end{equation}
The Dirac equations can be rewritten in terms of components of $\psi^i (z^i)$ as
	\begin{align}
	 	&\left[ \partial_{\bar{z}^i} + \frac{\pi q_a^A m_a^i }{2 {\rm Im} \tau^i} \left(z^i + \frac{q_a^A m_a^i \zeta_a^i}{q_a^A m_a^i}\right) \right] \psi_+^i (z^i,\bar{z}^i)=0, \\
	 	&\left[ \partial_{z^i} - \frac{\pi q_a^A m_a^i }{2 {\rm Im} \tau^i} \left(\bar{z}^i + \frac{q_a^A m_a^i \bar{\zeta}_a^i}{q_A^a m_a^i}\right) \right] \psi_-^i (z^i,\bar{z}^i)=0.
	\end{align}
Here, $\psi_+^i$ has degenerate zero-modes only if $M_A^i = q_a^A m_a^i > 0$, whereas $\psi_-^i$ has degenerate zero-modes only if $M_A^i<0$. Their degeneracy is given by $|M_A^i|$.
In addition, effective Wilson line $\zeta_A^i = \frac{q_a^A m_a^i \zeta_a^i}{q_a^A m_a^i}$ determines quasi-localization positions of wavefunctions of zero-modes.
Thus, Wilson lines are very important to Yukawa couplings.

If $M_A^i>0$, wavefunctions for $\psi_+^i$ are given by
\begin{equation}
	\psi_+^{i^{A,I}} = \Theta^{I,M_A^i} (z^i+\zeta_A^i,\tau_i),
\end{equation}
where
\begin{equation*}
	\begin{array}{ccl}
		\Theta^{I,M} (z,\tau) &=& {\cal N}_I \cdot e^{\pi i Mz {\rm Im}z/{\rm Im} \tau} \cdot \vartheta \left[ 
		\begin{array}{c}
			I/M \\
			0
		\end{array}
		\right] (Mz,M\tau),
		\\
		\vartheta \left[ 
		\begin{array}{c}
			a \\
			b
		\end{array}
		\right] (\nu,\tau) &=& \displaystyle\sum_{l \in {\bm Z}} e^{\pi i (a+l)^2 \tau} e^{2 \pi i (a+l) (\nu+b)},
	\end{array}	
\end{equation*}
and normalization factors ${\cal N}_I$ are determined such as 
\begin{equation}
		\int _{T^2} d^2 z \Theta^{I,M} (\Theta^{J,M})^* = \delta_{IJ}.
\end{equation}
The index $I = 0,\ldots,|M_A^i|$ labels degenerate zero-modes.
The total degeneracy, i.e. the number of generations is product of $|M_A^i|$,
\begin{equation}
	M_A = |M_A^{1}||M_A^{2}||M_A^{3}|.
\end{equation}

One can extract candidates for SM particles from adjoint representation of $SO(32)$ gauge group
with identification of hypercharge $U(1)_Y = (U(1)_3+3\sum_{a=4}^N U(1)_a)/6$,
where $N$ depends on models. ( See for detail \cite{Abe:2015mua}.)
These candidates are summarized as follows,
\begin{equation}
	\begin{array}{lll}
		Q : \left\{ 
			\begin{array}{l}
				Q_1 = (3,2)_{1,1,1;0,\dots,0} \\
				Q_2 = (3,2)_{-1,1,1;0,\dots,0}
			\end{array},
		\right. 
		&
		L : \left\{ 
			\begin{array}{l}
				L_1 = (1,2)_{1,1,-3;0,\dots,0} \\
				L_2 = (1,2)_{-1,1,-3;0,\dots,0}
			\end{array},
		\right. \\
		u_R : u_{R_2}^a = (3,1)_{0,1,1;\underline{1,0,\ldots,0}},
		&
		d_R : d_{R_3}^a = (3,1)_{0,1,1;\underline{-1,0,\ldots,0}}, \\
		e_R : u_{R_1}^a = (1,1)_{0,1,-3;\underline{-1,0,\ldots,0}},
		&
		\nu_R : n_2^a = (1,1)_{0,1,-3;\underline{1,0,\ldots,0}}, \\
		H_u : \bar{L}_4^a = (1,2)_{1,0,0;\underline{1,0,\ldots,0}} ,
		&
		H_d : L_3^a = (1,2)_{1,0,0;\underline{-1,0,\ldots,0}},
	\end{array}
\end{equation}
where indices imply $U(1)_{1,\ldots,13}$ charge $q_{1,\ldots,13}$ and 
the underlines are possible permutations.
Here, we focus on supersymmetric standard model, e.g., 
the minimal supersymmetric standard model (MSSM).
Here and hereafter, we use the superfield notation.
We can discuss non-supersymmetric SM similarly.

We need constraints on magnetic fluxes in order to make $U(1)_Y$ massless \cite{Abe:2015mua},
\begin{equation}
	\begin{array}{ccl}
		m_3^i &=& 0, \\
		m_{2+2a}^i &=& - m_{3+2a}^i \; (a=1,\ldots,\frac{N-3}{2}).
	\end{array}
\end{equation}
Furthermore, we impose K-theory constraints to construct models without heterotic five-branes,
\begin{equation}
	\sum_{a=1}^{2} m_a^i = 0 \; ({\rm mod} \; 2).
\end{equation}
We can achieve these conditions by setting
\begin{equation}
	\begin{array}{ll}
		M_{Q_2} = 3, & M_{L_2} = 3, \\
		M_{Q_1} = 0, & M_{L_1} = 0.
	\end{array}
\end{equation}
For the right-handed sector, we can obtain three generations of  quarks and leptons when $\sum_{a=4}^{13}M_{u_{R_2}^a} = -3$.
In general, there are many Higgs pairs, $H_u$ and $H_d$.

\subsection{Flavor symmetries in three generation models}

For the left-handed sector, three generations of quark and lepton doublets are realized by 12 cases, 
\begin{equation}
	M_{Q_2}^i = \left\{
	\begin{array}{l}
		(\underline{3,1,1}) \\
		(\underline{3,-1,-1}) \\
		(\underline{-3,-1,1}).
	\end{array}
	\right.
\label{eq:MQ2example}
\end{equation}
Since these cases are related with each other  by interchanging two tori $(T^2)_i \leftrightarrow (T^2)_j$, or changing signs of magnetic fluxes on two tori $m_a^i \rightarrow -m_a^i,m_a^{j} \rightarrow -m_a^{j}$, we can set 
\begin{equation}
	M_{Q_2}^i = (-3,-1,1),
\label{eq:MQ2}
\end{equation}
without losing generality.

For the right-handed sector, we have a lot of models to realize three generations of quarks and leptons. The first example is 
obtained as follows,  
\begin{equation}
	\begin{array}{lll}
		M_{u_{R_2}^4}^i = M_{u_{R_2}^6}^i = M_{u_{R_2}^8}^i, & M_{u_{R_2}^4} =-1, & \sum_{a=4}^{13} M_{u_{R_2}^a} = -3
	\end{array}.
\end{equation}
In this model, the gauge symmetries enhance to larger one, 
$\prod_{a=4}^9 U(1)_a \rightarrow SU(3)_u \times SU(3)_d \times SU(2)_R$.
Cartan elements of $SU(3)_u$ are $H_4-H_6,H_4+H_6-2H_8$.
$SU(3)_d$ and $SU(2)_R$ are given by $H_5-H_7,H_5+H_7-2H_9$ and $H_4+H_6+H_8-H_5-H_7-H_9$, respectively.
These $SU(3)_{u,d}$ symmetries are flavor symmetries among the right-handed quarks and leptons 
as well as Higgs fields.
That is, the right-handed quarks in the up-sector (down-sector) are a triplet under  $SU(3)_{u}$ ($SU(3)_{d}$ ).
Similarly, the Higgs fields $H_u$ ($H_d$) are also triplets under  $SU(3)_{u}$ ($SU(3)_{d}$ ), 
while the right-handed neutrinos (charged leptons) are a triplet under  $SU(3)_{u}$ ($SU(3)_{d}$ ).
Thus, we refer to this model as the $SU(3)_f$ model.
The left-handed quarks and leptons are singlets under $SU(3)_{u,d}$ symmetries.

The second example is obtained as 
\begin{equation}
	\begin{array}{ll}
		M_{u_{R_2}^4}^i=-M_{Q_2}^i, & \sum_{a=5}^{13}M_{u_{R_2}^a} = 0
	\end{array}.
\end{equation}
This model has the gauge symmetry $SU(2)_R$, whose Cartan element is $H_4-H_5$.
In addition, this model has non-Abelian discrete symmetry $\Delta(27)$ \cite{Abe:2009vi}.
The three generations of the quarks and leptons are triplets under $\Delta(27)$.
The Higgs fields are also $\Delta(27)$ triplets.

There are other models, which have different flavor structures.
We focus on the above two models, the $SU(3)_f$ flavor model and 
the $\Delta(27)$ flavor model, since they contain good flavor symmetries, leading simple mass matrices. 
Throughout this paper, we also assume that the gauge couplings of those flavor symmetries are 
enough suppressed at the low-energy scale, although it depends on the matter contents of hidden sector.

\subsection{Computation of Yukawa couplings}

As shown in the previous section, the wavefunction of each degenerate mode on tori is quasi-localized at a different point, 
which is controlled by Wilson lines.
Since performing overlap integral derives Yukawa couplings, those couplings can become hierarchical.
Let us now compute Yukawa couplings. 
Yukawa coupling in 4D is given by product of three overlap integrals on three 2-tori, i.e.
\begin{equation}
	\begin{array}{ccl}
		Y_{{\cal IJK}} &=& g \lambda^{(1)}_{I_1 J_1 K_1} \lambda^{(2)}_{I_2 J_2 K_2} \lambda^{(3)}_{I_3 J_3 K_3},\\
		\lambda^{(i)}_{I_i J_i K_i} &=& \int_{(T^2)_i} d^2 z^i \; \Theta^{I_i,M_A^i} (z^i + \zeta_A^i,\tau_i) \Theta^{J_i,M_B^i} (z^i + \zeta_B^i,\tau_i) \left( \Theta^{K_i,-M_C^i} (z^i + \zeta_C^i,\tau_i) \right)^*,
	\end{array}
\label{eq:Yukawa coupling}
\end{equation}
where $g$ is the 4D gauge coupling, ${\cal I}= (I_1,I_2,I_3), {\cal J} = (J_1,J_2,J_3), {\cal K} = (K_1,K_2,K_3)$, and 
we impose invariance under $U(1)_a$ gauge symmetries, $q_a^A+q_a^B+q_a^C=0$.
Note that the Lorentz symmetry of the 6D compact space also leads to the 
selection rule of allowed Yukawa couplings.
For example, the Yukawa coupling, $Y^{(u)}H_uQ_Lu_R$,  is allowed only if 
the fermionic compoenets of $H_u$, $Q_L$ and $u_R$ have the chiralities,  
$(+,-,-)$, $(-,+,-)$ and $(-,-,+)$ in the 6D compact space, respectively, 
and other permutations.

By performing overlap integral, we obtain
\begin{equation}
	\begin{array}{ccl}
		\lambda_{I_i J_i K_i} &=& \frac{ {\cal N}_{I_i} {\cal N}_{J_i} }{ {\cal N}_{K_i} } e^{\pi i (M_A^i \zeta_A^i {\rm Im} \zeta_A^i + M_B^i \zeta_B^i {\rm Im} \zeta_B^i + M_C^i \zeta_C^i {\rm Im} \zeta_C^i ) / {\rm Im} \tau^i} \\
		&& \cdot \sum_{m \in {\bm Z}_{M_A^i+M_B^i}} \vartheta \left[
			\begin{array}{c}
				\frac{M_B^i I_i-M_A^i J_i+M_A^i M_B^i m}{M_A^i M_B^i (-M_C^i)} \\
				0
			\end{array}
			\right] (M_A^i M_B^i (\zeta_A^i-\zeta_B^i),\tau M_A^i M_B^i (-M_C^i)) \\
		&& \cdot \delta_{I_i+J_i+M_A^i m,K_i}.
	\end{array}
	\label{eq:Yukawa}
\end{equation}


\section{Quark masses and mixings}
In this section, we study the mass matrices and mixing angles of quark sector.

\subsection{$SU(3)_f$ model}


We begin with the $SU(3)_f$ model.
Although there are several $SU(3)_f$ models, 
we focus on the case $M_{{u_R^c}^4_2}^i=(-1,1,-1)$ such that the Lorentz symmetry of the 6D compact space allows Yukawa couplings.
The three generations of the up-sector (down-sector) right-handed quarks are a triplet under 
$SU(3)_u$ ($SU(3)_d$).
This model contains totally $(4 \times 3)$ pairs of vector-like Higgs fields, 
and these up-sector (down-sector) Higgs fields are 4 triplets under $SU(3)_u$ ($SU(3)_d$).
The degeneracy factor, 4, comes from 4 chiral zero-modes on the first $T^2$.
For simplicity, we concentrate ourselves on a single zero-mode among $4$ zero-modes 
in order to study the properties of $SU(3)_f$ flavor model.
Note that the difference among 4 chiral zero-modes on the first $T^2$ is the peak positions of wave-functions, 
and the peak position can be shifted by varying the Wilson line.
That implies that any choice of a single zero-mode among $4$ zero-modes can lead to equivalent configuration 
by varying Wilson lines.
Thus, we consider 3 pairs of Higgs fields, which are triplets under $SU(3)_u$ and $SU(3)_d$, 
and we denote them by $H_{uK}$ and $H_{dK}$ with $K=0,1,2$.

Yukawa coupling terms of the up-sector quarks and 3 Higgs fields, 
\begin{equation}
	Y^{(u)}_{IJK} H_{u K} Q_{L_I} u_{R_J},
\end{equation}
can be written by

\begin{equation}
	\begin{array}{c}
		{Y^{(u)}_{IJ0}} =g \begin{pmatrix}
			\eta_{8,\zeta_{u1}} & 0 & 0 \\
			\eta_{4,\zeta_{u1}} & 0 & 0 \\
			\eta_{0,\zeta_{u1}} & 0 & 0
		\end{pmatrix}, \\
		{Y^{(u)}_{IJ1}}  = g \begin{pmatrix}
			0 & \eta_{8,\zeta_{u2}} & 0 \\
			0 & \eta_{4,\zeta_{u2}} & 0 \\
			0 & \eta_{0,\zeta_{u2}} & 0
		\end{pmatrix}, \\
		{Y^{(u)}_{IJ2}}  =g \begin{pmatrix}
			0 & 0 & \eta_{8,\zeta_{u3}} \\
			0 & 0 & \eta_{4,\zeta_{u3}} \\
			0 & 0 & \eta_{0,\zeta_{u3}}
		\end{pmatrix},
	\end{array}
\end{equation}
up to the normalization factors,
where $\eta_{n,\zeta_{ui}}$ is contributions on Yukawa couplings from 
the first $T^2$, and is obtained by use of Eq.~(\ref{eq:Yukawa}).
In the following analysis, we restrict complex structure moduli $\tau_i$ and Wilson lines $\zeta_a^i$ are pure imaginary.
Then, $\eta_{n,\zeta_{ui}}$ is written by 
\begin{eqnarray}
& & 	\eta_{n,\zeta_{ui}} = \sum_l e^{- 12\pi {\rm Im} \tau (\frac{n}{12}+l+\frac{ {\rm Im} \zeta_{ui} }{ {\rm Im} \tau_{1}})^2},
\end{eqnarray}
where
\begin{eqnarray}
	& & \zeta_{ui} = (m_2^{1}+m_{2i+2}^{1}) m_1^{1} \zeta_1^{1}-(m_1^{1}-m_{2i+2}^{1}) m_2^{1} \zeta_2^{1}-(m_1^{1}+m_2^{1}) m_{2i+2}^{1} \zeta_{2i+2}^{1}.
\end{eqnarray}
We obtain $\eta_{0,\zeta_{ui}} \sim 1$ for $\zeta_{ui} =0$.

Similarly, the down sector Yukawa couplings are written in the same form except replacing 
$\eta_{n,\zeta_{ui}}$ by $\eta_{n,\zeta_{di}}$.
Wilson lines for the down sector are defined by
\begin{eqnarray}
	& & \zeta_{di} = (m_2^{1}+m_{2i+3}^{1}) m_1^{1} \zeta_1^{1}-(m_1^{1}-m_{2i+3}^{1}) m_2^{1} \zeta_2^{1}-(m_1^{1}+m_2^{1}) m_{2i+3}^{1} \zeta_{2i+3}^{1}.
\end{eqnarray}

Here, we assume that these Higgs fields develop their vacuum expectation values (VEVs).
That leads to the following mass matrix for the up-sector
\begin{equation}
	M^u = 
g \langle H_{u2} \rangle
	\begin{pmatrix}
		\eta_{8,\zeta_{u_1}} \rho_{u1} & \eta_{8,\zeta_{u_2}} \rho_{u2} & \eta_{8,\zeta_{u_3}} \\
		\eta_{4,\zeta_{u_1}} \rho_{u1} & \eta_{4,\zeta_{u_2}} \rho_{u2} & \eta_{4,\zeta_{u_3}} \\
		\eta_{0,\zeta_{u_1}} \rho_{u1} & \eta_{0,\zeta_{u_2}} \rho_{u2} & \eta_{0,\zeta_{u_3}}
	\end{pmatrix},
\label{eq:massmatrixsu3}
\end{equation}
and the down sector mass matrix
\begin{equation}
	M^d = 
g \langle H_{d2} \rangle
	\begin{pmatrix}
		\eta_{8,\zeta_{d_1}} \rho_{d1} & \eta_{8,\zeta_{d_2}} \rho_{d2} & \eta_{8,\zeta_{d_3}} \\
		\eta_{4,\zeta_{d_1}} \rho_{d1} & \eta_{4,\zeta_{d_2}} \rho_{d2} & \eta_{4,\zeta_{d_3}} \\
		\eta_{0,\zeta_{d_1}} \rho_{d1} & \eta_{0,\zeta_{d_2}} \rho_{d2} & \eta_{0,\zeta_{d_3}}
	\end{pmatrix},
\label{eq:massmatrixsd3}
\end{equation}
where  
\begin{equation}
\rho_{u1} = \frac{\langle H_{u0} \rangle }{\langle H_{u2}\rangle }, \qquad \rho_{u2} =  \frac{\langle H_{u1} \rangle }{\langle H_{u2}\rangle },
\end{equation}
\begin{equation}
\rho_{d1} = \frac{\langle H_{d0} \rangle }{\langle H_{d2}\rangle }, \qquad \rho_{d2} =  \frac{\langle H_{d1} \rangle }{\langle H_{d2}\rangle }.
\end{equation}
The mass ratios and mixing anlges are determined by the complex structure $\tau_1$ on the first $T^2$, 
Wilson lines $\zeta_{ui}$ and $\zeta_{di}$ and ratios $\rho_{u1}, \rho_{u2}, \rho_{d1}, \rho_{d2}$.
In this paper, we treat them as free parameters to fit the data, although 
they are determined by the stabilization of moduli and Higgs fields.

The above matrices for up-sector have the hierarchy, $M^{u}_{ij} \le M^{u}_{i'j'}$
for $i \le i'$ and $j \le j'$ when $\zeta_{u1} \sim \zeta_{u2} \sim \zeta_{u3} \sim 0$. 
Down-sector matrices have same characteristics.

Let us consider the $(2 \times 2)$ lower right submatrix first.
Because of the hierarchical structure, 
the diagonalizing angles of the up-and down-sector mass matrices are estimated as 
\begin{equation}
	\theta_{23}^{u,d} \sim M_{23}^{u,d}/M_{33}^{u,d},
\end{equation}
and the mass ratios are also estimated as 
\begin{equation}
	(m_2/m_3)^{u,d} \sim |M_{22}^{u,d}/M_{33}^{u,d}-(M_{23}^{u,d}/M_{33}^{u,d})(M_{32}^{u,d}/M_{33}^{u,d})|.
\end{equation}
Similarly, we can examine the  $(2 \times 2)$ upper left submatrix to estimate 
 diagonalizing angles $\theta_{12}^{u,d}$ and  $\theta_{13}^{u,d}$ as well as mass ratios.
Then, the Cabibbo-Kobayashi-Masukawa(CKM) matrix,
\begin{equation}
	V_{CKM} = 
	\begin{pmatrix}
		V_{ud} & V_{us} & V_{ub} \\
		V_{cd} & V_{cs} & V_{cb} \\
		V_{td} & V_{ts} & V_{tb}
	\end{pmatrix},
\end{equation}
is estimated as 
\begin{equation}
	\begin{array}{ccl}
		|V_{us}| & \sim & |\theta_{12}^u-\theta_{12}^d|, \\
		|V_{ub}| & \sim & |\theta_{13}^u-\theta_{13}^d|, \\
		|V_{cb}| & \sim & |\theta_{23}^u-\theta_{23}^d|.
	\end{array}
\end{equation}
These experimental values are 
\begin{equation}
	\begin{array}{ccl}
		|V_{us}| & = & 0.23, \\
		|V_{ub}| & = & 0.0041, \\
		|V_{cb}| & = & 0.041.
	\end{array}
\end{equation}
The renormalization group flow in the SM leads
\begin{equation}
	\begin{array}{ccl}
		m_u / m_t &\sim& 6.5 \times 10^{-6}, \\
		m_c / m_t &\sim& 3.2 \times 10^{-3}, \\
		m_d / m_b &\sim& 1.1 \times 10^{-3}, \\
		m_s / m_b &\sim& 2.2 \times 10^{-2},
	\end{array}
\end{equation}
at $\Lambda_{{\rm GUT}}=2 \times 10^{16}$GeV. ( see e.g.\cite{Xing:2007fb}.)
The RG flow of the MSSM also leads to similar values.


With hierarchical Yukawa matrices, we can estimate mass ratios and mixing angles for up-sector,
\begin{equation}
	\begin{array}{ccl}
		(m_1/m_3)^{u} & \sim & \rho_{u1} |\frac{\eta_{8,\zeta_{u1}}}{\eta_{0,\zeta_{u3}}} - \frac{\rho_{u2}}{(m_2/m_3)^{u}} \frac{\eta_{4,\zeta_{u1}}}{\eta_{0,\zeta_{u3}}} \frac{\eta_{8,\zeta_{u2}}}{\eta_{0,\zeta_{u3}}}|, \\
		(m_2/m_3)^{u} & \sim & \rho_{u2}|\frac{\eta_{4,\zeta_{u2}}}{\eta_{0,\zeta_{u3}}} - \frac{\eta_{4,\zeta_{u3}}}{\eta_{0,\zeta_{u3}}} \frac{\eta_{0,\zeta_{u2}}}{\eta_{0,\zeta_{u3}}}|, \\
		m_3^{u} & \sim & g \langle H_{u2} \rangle \eta_{0,\zeta_{u3}}\\
		\theta_{12}^{u} & \sim & \frac{\rho_{u2}}{(m_2/m_3)^{u}}\frac{\eta_{8,\zeta_{u2}}}{\eta_{0,\zeta_{u3}}}, \\
		\theta_{13}^{u} & \sim & \frac{\eta_{8,\zeta_{u3}}}{\eta_{0,\zeta_{u3}}}, \\
		\theta_{23}^{u} & \sim & \frac{\eta_{4,\zeta_{u3}}}{\eta_{0,\zeta_{u3}}}.
	\end{array}
\label{eq:estimation}
\end{equation}
Down-sector gives similar expressions.

When $\rho_{ui} \sim \rho_{di} \sim 1$, the ratios of the above parameters 
 bring insufficient hierarchy to realize the mixing angles, 
thus we need {\it tuning} to realize hierarchical structure.
Here we show an example of set of parameters, yielding realistic quark masses and mixings.
We set
\begin{equation}
	\begin{array}{lll}
		\tau_1 &=& 1.1i, \\
		\zeta_{u_i} &=& (-0.065i, -0.068i, -0.072i), \\
		\zeta_{d_i} &=& (0.002i, -0.063i, 0.017i), \\
		\rho_{u_i} &=& (1, 1), \\
		\rho_{d_i} &=& (1, 1).
	\end{array}
\label{eq:exampleparameterssu3-1}
\end{equation}
Note that this model contains {\it tuning}.
For instance, $(m_2/m_3)^{u}$ is estimated as $|0.056-0.061|=0.005$ in Eq.(\ref{eq:estimation}), indicating cancellation derives hierarchical mass ratio.
Similar cancellation is required to derive other mass ratios.
Since $\rho_{ui}, \rho_{di} \sim $ ${\cal O}(1)$ do not suppress mass ratio, we need {\it tuning} to realize hierarchical masses.
These parameters lead to realistic values shown in Table \ref{table:numericalexamplesu3-1}.


\begin{table}[h]
	\begin{center}
		\begin{tabular}{|c||c|} \hline
			$(m_u/m_t,m_c/m_t)$ & $(6.3 \times 10^{-6}, 4.0 \times 10^{-3})$ \\ \hline
			$(m_d/m_b,m_s/m_b)$ & $(1.6 \times 10^{-3}, 1.9 \times 10^{-2})$ \\ \hline \hline
			$|V_{{\rm CKM}}|$ & $\begin{pmatrix}
	 			0.97 & 0.23 & 0.012 \\
	 			0.23 & 0.97 & 0.039 \\
				0.021 & 0.035 & 1.0
			\end{pmatrix}$ \\ \hline
		\end{tabular}
	\end{center}
\caption{Mass ratios and mixings evaluated with values of complex structure moduli on first $T^2$, Higgs VEVs and Wilson lines in Eq. (\ref{eq:exampleparameterssu3-1}).}
\label{table:numericalexamplesu3-1}
\end{table}


When $\rho_{ui}$, $\rho_{di}$ are not of ${\cal O}(1)$, but hierarchical, we do not need tuning.
Next, we show an example without {\it tuning}. We set 
\begin{equation}
	\begin{array}{lll}
		\tau_1 &=& 1.1i, \\
		\zeta_{u_i} &=& (0.010i, -0.035i, -0.020i), \\
		\zeta_{d_i} &=& (-0.020i, -0.084i, -0.070i), \\
		\rho_{u_i} &=& (0.0021, 0.44), \\
		\rho_{d_1} &=& (0.18, 0.97).
	\end{array}
\label{eq:exampleparameterssu3-2}
\end{equation}
leading to result shown in table \ref{table:numericalexamplesu3-2}.


\begin{table}[h]
	\begin{center}
		\begin{tabular}{|c||c|} \hline
			$(m_u/m_t,m_c/m_t)$ & $(8.7 \times 10^{-6}, 2.8 \times 10^{-3})$ \\ \hline
			$(m_d/m_b,m_s/m_b)$ & $(4.4 \times 10^{-4}, 1.4 \times 10^{-2})$ \\ \hline \hline
			$|V_{{\rm CKM}}|$ & $\begin{pmatrix}
	 			0.98 & 0.20 & 0.018 \\
	 			0.20 & 0.98 & 0.049 \\
	 			0.0076 & 0.051 & 1.0
			\end{pmatrix}$ \\ \hline
		\end{tabular}
	\end{center}
\caption{Mass ratios and mixings evaluated with values of complex structure moduli on first $T^2$, Higgs VEVs and Wilson lines in Eq. (\ref{eq:exampleparameterssu3-2}).}
\label{table:numericalexamplesu3-2}
\end{table}


\subsection{{$\Delta(27)$ model}}



Let us move on to the $\Delta (27)$ flavor symmetry model.
In this model, all of quarks and leptons are the same type of triplets of $\Delta (27)$.\footnote{
There are several types of triplets in $\Delta(27)$ \cite{Ishimori:2010au}.}
We focus on the case $M_{{u_R^c}^4_2}^i=(-3,1,-1)$ to obtain full-rank mass matrices.
This model contains ($6=2 \times 3 $) pairs of vector-like Higgs fields, 
and they are 2 triplets of $\Delta (27)$, which are also the same type of triplets as quarks and leptons.
The degeneracy factor, 6, comes from 6 chiral zero-modes on first $T^2$.

We use all pairs of Higgs fields to realize realistic mass matrices, 
which are two triplets under $\Delta (27)$. 
We denote them by $H_{uK}$ and $H_{dK}$ with $K=0,\ldots,5$.
Among them  $H_{uK}$ as well as $H_{dK}$ with $K=0,1,2$ correspond to a triplet, 
while  $H_{uK}$ as well as $H_{dK}$ with $K=3,4,5$ correspond to another triplet,
They  lead to Yukawa coupling term

\begin{equation}
	Y^{(u)}_{IJK} H_{u K} Q_{L_I} u_{R_J},
\end{equation}
which can be written by 
\begin{equation}
	\begin{array}{ccc}
		Y_{IJ0}^{(u)} = g \begin{pmatrix}
			\tilde{\eta}_{0,\zeta_{u}} & 0 & 0 \\
			0 & 0 & \tilde{\eta}_{6,\zeta_{u}} \\
			0 & \tilde{\eta}_{12,\zeta_{u}} & 0
		\end{pmatrix}, 
		&
		Y_{IJ1}^{(u)} = g \begin{pmatrix}
			0 & \tilde{\eta}_{15,\zeta_{u}} & 0 \\
			\tilde{\eta}_{3,\zeta_{u}} & 0 & 0 \\
			0 & 0 & \tilde{\eta}_{9,\zeta_{u}}
		\end{pmatrix}, \\ 
		\\
		Y_{IJ2}^{(u)} = g \begin{pmatrix}
			0 & 0 & \tilde{\eta}_{12,\zeta_{u}} \\
			0 & \tilde{\eta}_{0,\zeta_{u}} & 0 \\
			\tilde{\eta}_{6,\zeta_{u}} & 0 & 0
		\end{pmatrix}, 
		&
		Y_{IJ3}^{(u)} = g \begin{pmatrix}
			\tilde{\eta}_{9,\zeta_{u}} & 0 & 0 \\
			0 & 0 & \tilde{\eta}_{15,\zeta_{u}} \\
			0 & \tilde{\eta}_{3,\zeta_{u}} & 0
		\end{pmatrix}, \\
		\\
		Y_{IJ4}^{(u)} = g \begin{pmatrix}
			0 & \tilde{\eta}_{6,\zeta_{u}} & 0 \\
			\tilde{\eta}_{12,\zeta_{u}} & 0 & 0 \\
			0 & 0 & \tilde{\eta}_{0,\zeta_{u}}
		\end{pmatrix},
		&
		Y_{IJ5}^{(u)} = g \begin{pmatrix}
			0 & 0 & \tilde{\eta}_{3,\zeta_{u}} \\
			0 & \tilde{\eta}_{9,\zeta_{u}} & 0 \\
			\tilde{\eta}_{15,\zeta_{u}} & 0 & 0
		\end{pmatrix},
	\end{array}
\end{equation}
up to the normalization factors,
where $\tilde{\eta}_{n,\zeta_u}$ is contributions on Yukawa couplings from 
the first $T^2$, again.
As $SU(3)_f$ model, we restrict that complex structure moduli $\tau$ and Wilson lines $\zeta_a$ are pure imaginary.
Then $\tilde{\eta}_{n,\zeta_u}$ is written by 
\begin{eqnarray}
& & \tilde{\eta}_{n,\zeta_u} = \sum_l \sum_{m=0}^2 e^{- 54\pi {\rm Im} \tau (\frac{n}{54}+\frac{m}{3}+l+\frac{ {\rm Im} \zeta_u }{ {\rm Im} \tau_{1}})^2},
\end{eqnarray}

Similarly, the down sector Yukawa couplings are written in the same form except replacing 
$\tilde{\eta}_{n,\zeta_u}$ by 
$\tilde{\eta}_{n,\zeta_d}$.
Wilson lines for the up and down-sectors are
\begin{eqnarray}
	& & \zeta_{u} = (m_2^{1}+m_{4}^{1}) m_1^{1} \zeta_1^{1}-(m_1^{1}-m_{4}^{1}) m_2^{1} \zeta_2^{1}-(m_1^{1}+m_2^{1}) m_{4}^{1} \zeta_{4}^{1},
\end{eqnarray}
and
\begin{eqnarray}
	& & \zeta_{d} = (m_2^{1}+m_{5}^{1}) m_1^{1} \zeta_1^{1}-(m_1^{1}-m_{5}^{1}) m_2^{1} \zeta_2^{1}-(m_1^{1}+m_2^{1}) m_{5}^{1} \zeta_{5}^{1}.
\end{eqnarray}

Note that $Y_{IJm}$, ($m=0,1,2$), have hierarchy opposite to $Y_{IJm+3}$, not preferred to realize hierarchical Yukawa matrix.
We assume that $H_{u2}$, $H_{u3}$ and $H_{u4}$ develop their VEVs.
Then, the mass matrix of  up-sector quarks is obtained as ,
\begin{equation}
	M^{u} \approx g \langle H_{u4} \rangle
	\begin{pmatrix}
		\tilde{\eta}_{9,\zeta_u} \rho_{u3} & \tilde{\eta}_{6,\zeta_u} & \tilde{\eta}_{12,\zeta_u} \rho_{u2} \\
		\tilde{\eta}_{12,\zeta_u} & \tilde{\eta}_{0,\zeta_u} \rho_{u2} & \tilde{\eta}_{15,\zeta_u} \rho_{u3} \\
		\rho_{u2} \tilde{\eta}_{6,\zeta_u} & \tilde{\eta}_{3,\zeta_u} \rho_{u3} & \tilde{\eta}_{0,\zeta_u}
	\end{pmatrix},
\end{equation}
where $\rho_{ui} = \frac{\langle H_{ui} \rangle }{\langle H_{u4} \rangle }$ with $i=2,3$.
For the down-sector, $\rho_{d3} \tilde{\eta}_{9,\zeta_d}$ is too small to realize down quark mass.
Thus, we assume that $H_{d0} $ as well as $H_{d2}$, $H_{d3}$ and $H_{d4}$ develop 
their VEVs.
Then, the mass matrix of the down-sector quarks is given by
\begin{equation}
	M^{d} \approx g \langle H_{d4} \rangle
	\begin{pmatrix}
		\tilde{\eta}_{0,\zeta_d} \rho_{d0} & \tilde{\eta}_{6,\zeta_d} & \tilde{\eta}_{12,\zeta_d} \rho_{d2} \\
		\tilde{\eta}_{12,\zeta_d} & \tilde{\eta}_{0,\zeta_d} \rho_{d2} & \tilde{\eta}_{15,\zeta_d} \rho_{d3} \\
		\tilde{\eta}_{6,\zeta_d} \rho_{d2} & \tilde{\eta}_{3,\zeta_d} \rho_{d3} & \tilde{\eta}_{0,\zeta_d}
	\end{pmatrix},
\end{equation} 
where $\rho_{di} = \frac{\langle H_{di} \rangle }{\langle H_{d4} \rangle }$ with $i=0,2,3$.

Since $(m_u/m_t)(m_c/m_t) = {\rm det} (M^u)/(m_t)^3 \sim {\rm det} (Y_{IJ4}/\tilde{\eta}_{0,\xi_u})$ 
leads to constraint on ${\rm Im} \tau_1$,
$(\tilde{\eta}_{6,\zeta_u})(\tilde{\eta}_{12,\zeta_u}) \sim e^{-\frac{4}{3} \pi {\rm Im} \tau_1} \approx 2 \times 10^{-8}$, 
we set ${\rm Im} \tau_1 = 4.2$.
Next, we concentrate on $2 \times 2$ low right matrices,
\begin{equation}
	v_4^{u,d}
	\begin{pmatrix}
		\rho_{u,d2} \tilde{\eta}_{0,\zeta_{u,d}} & \rho_{u,d3} \tilde{\eta}_{15,\zeta_{u,d}} \\
		\rho_{u,d3} \tilde{\eta}_{3,\zeta_{u,d}} & \tilde{\eta}_{0,\zeta_{u,d}}
	\end{pmatrix},
\end{equation}
leading
\begin{equation}
	\begin{array}{ccl}
		V_{cb} & \sim & \rho_{u3} \tilde{\eta}_{15,\zeta_u}/ \tilde{\eta}_{0,\zeta_u}-\rho_{d3} \tilde{\eta}_{15,\zeta_d}/ \tilde{\eta}_{0,\zeta_d}, \\
		m_c/m_t & \sim & \rho_{u2} - (\rho_{u3})^2 \tilde{\eta}_{3,\zeta_u} \tilde{\eta}_{15,\zeta_u}/(\tilde{\eta}_{0.\zeta_u})^2, \\
		m_s/m_b & \sim & \rho_{d2} - (\rho_{d3})^2 \tilde{\eta}_{3,\zeta_d} \tilde{\eta}_{15,\zeta_d}/(\tilde{\eta}_{0,\zeta_d})^2.
	\end{array}
\end{equation}
Then, we can estimate $\rho_{u2} \sim 3.2 \times 10^{-3},\rho_{d2} \sim 2.2 \times 10^{-2},\rho_{u3}-\rho_{d3} \sim \pm 0.36$,
assuming $\zeta_u = \zeta_d = 0$.
Finally, we use $Y_0$ to realize $m_d$.
In a way similar to up-sector mass matrix, we set $\rho_{d0} \sim 1.1 \times 10^{-3}$ from the constraint $\det (M^d) \sim \rho_{d0} \rho_{d2} \tilde{\eta}_{0,\zeta_d}^3$. 
In the following representative parameters,

\begin{equation}
	\begin{array}{ccl}
		\tau & = & 4.2i, \\
		\zeta_u & = & 0.0045i, \\
		\zeta_d & = & -0.1i, \\
		\rho_{ui} & = & (0, 0, 0.0053, 0.415, 1, 0), \\
		\rho_{di} & = & (0.0012, 0, 0.027, 0.56, 1, 0),
	\end{array}
\label{eq:exampleparametersdelta27}
\end{equation}
we obtain the realistic quark masses and mixings shown in Table \ref{table:numericalexampledelta27}.


\begin{table}[h]
	\begin{center}
		\begin{tabular}{|c||c|} \hline
			$(m_u/m_t,m_c/m_t)$ & $(7.2 \times 10^{-6}, 3.2 \times 10^{-3})$ \\ \hline
			$(m_d/m_b,m_s/m_b)$ & $(1.1 \times 10^{-3}, 2.1 \times 10^{-2})$ \\ \hline \hline
			$|V_{{\rm CKM}}|$ & $\begin{pmatrix}
	 			0.97 & 0.23 & 0.0019 \\
				0.23 & 0.97 & 0.033 \\
				0.0095 & 0.031 & 1.0
	\end{pmatrix}$ \\ \hline
		\end{tabular}
	\end{center}
\caption{Mass ratios and mixings evaluated with values of complex structure moduli on first $T^2$, Higgs VEVs and Wilson lines in Eq. (\ref{eq:exampleparametersdelta27}).}
\label{table:numericalexampledelta27}
\end{table}

\newpage

\section{Lepton sector}

Here, we give comments on the lepton sector.

As mentioned in section 2.1, when magnetic flux and Wilson lines along the $U(1)_3$ direction 
are vanishing, the $SU(3)_C$ gauge symmetry is enhanced to $SU(4)$.
In such a case, the charged lepton mass matrix is the same as 
the down-sector quark mass matrix.
Let us consider the model, where this $SU(4)$ is broken only by Wilson lines.
That is, we introduce different Wilson lines between the down-sector quarks and charged lepton sectors.
Then, the charged lepton mass matrix corresponding to section 3.1 can be written,
\begin{equation}
	M^l = 
g \langle H_{d2} \rangle
	\begin{pmatrix}
		\eta_{8,\zeta_{l_1}} \rho_{d1} & \eta_{8,\zeta_{l_2}} \rho_{d2} & \eta_{8,\zeta_{l_3}} \\
		\eta_{4,\zeta_{l_1}} \rho_{d1} & \eta_{4,\zeta_{l_2}} \rho_{d2} & \eta_{4,\zeta_{l_3}} \\
		\eta_{0,\zeta_{l_1}} \rho_{d1} & \eta_{0,\zeta_{l_2}} \rho_{d2} & \eta_{0,\zeta_{l_3}}
	\end{pmatrix},
\label{eq:massmatrixsl3}
\end{equation}
for the $SU(3)_f$ model.
Here, the new parameters in the lepton sector are the Wilson lines, $\zeta_{l_i}$. 
The experimental values of mass ratios in the charged lepton sector, $m_e/m_\tau$ and $m_\mu/m_\tau$, 
are similar to those in the down-sector quarks, $m_d/m_b$ and $m_s/m_b$.
Thus, we can realize the charged lepton mass ratios by setting $\zeta_{l_i} \sim \zeta_{d_i} $.
Similarly, we can discuss the charged lepton sector for the $\Delta(27)$ model.
Thus, it is straightforward to realize the charged lepton mass ratios in 
both the $SU(3)_f$ model and $\Delta(27)$ model.

We may assign the right-handed neutrinos such that they can 
couple with the left-handed leptons and up-sector Higgs scalars.
That is the assignment in section 2.
Then, in order to discuss the neutrino masses, we need to study the origin 
of right-handed Majorana masses.
Our models do not include singlets, whose VEVs become right-handed Majorana mass terms 
in the 3-point couplings, because of gauge invariances of extra $U(1)$ symmetries.
Thus, right-handed Majorana mass terms would be generated by higher dimensional terms 
or non-perturbative terms.
Such non-perturbative terms may be constrained by extra anomalous $U(1)$ symmetries, 
because factors in non-perturbative terms, $e^{-aS-b_iT_i}$, have anomalous $U(1)$ charges.

In the $SU(3)_f$ model, the three generations of neutrinos in the above assignment 
correspond to a $SU(3)_u$ triplet and they have the same extra $U(1)$ charge.
Thus, their Majorana mass terms can not be generated unless the $SU(3)_u$ symmetry is broken.
On the other hand, once the $SU(3)_u$ symmetry is broken, such mass terms would be generated 
but its pattern depends on the breaking pattern.
For example, it is possible to break $SU(3)_u$ such that breaking does not induce a large mass ratio 
among the triplets and Majorana mass terms realize large mixing angles.

In the $\Delta(27)$ model, three generations of right-handed neutrinos 
are $\Delta(27)$ triplets.
Again,  unless the $\Delta(27)$ symmetry is broken, 
their Majorana mass terms are not generated.
On the hand, non-perturbative effects may break the  $\Delta(27)$ symmetry.\footnote{See for anomalies of 
non-Abelian discrete symmetries \cite{Araki:2008ek,Hamada:2014hpa}.}
In such a case, all of entries may be allowed.
Because three generations of right-handed neutrinos have the same 
extra $U(1)$ charges, those entries in the Majorana mass would be of 
the same order, and we may have large mixing angles.

Also, we can comment on the Higgs $\mu$-term matrix.
Our models have no singlets $S$, which have perturbative 3-point couplings with 
the Higgs pairs, $SH_u H_d$ like the next-to-minimal supersymmetric standard model, 
because extra $U(1)$ symmetries forbid such couplings.
Higher order couplings or non-perturbative effects would generate the $\mu$-terms. 
In the  $SU(3)_f$ model,  $H_u$ and $H_d$ are triplets 
under $SU(3)_u$ and $SU(3)_d$, respectively.
Thus, unless those symmetries are broken, 
$\mu$-terms can not be generated.
Similar to the above comment on the neutrino masses, the pattern of 
the $\mu$-term matrix depends on their breaking.
It is plausible that the triplets develop 
similar  VEVs such as $\langle H_{u0} \rangle \sim \langle H_{u1} \rangle \sim \langle H_{u2} \rangle $, 
and $\langle H_{d0} \rangle = \langle H_{d1} \rangle =\langle H_{d2} \rangle $.
The situation of the $\mu$-term in the $\Delta(27)$ is similar.

\section{Conclusion}

We have studied quark mass matrices in $SO(32)$ heterotic string theory 
on 6D torus with magnetic fluxes.
We have examined two models, the $SU(3)_f$ flavor model and the $\Delta(27)$ model.
In both models, we have realized realistic quark masses and mixing angles 
by using our parameters, the complex structure, Wilson lines as well as 
Higgs VEV ratios.
Similarly, we can discuss the charged lepton masses.

We have used the complex structure and Wilson lines as free parameters.
It is important to discuss dynamics to determine those values.
That is beyond our scope.

Our models do not have Majorana right-handed neutrino mass terms at tree-level 
or singlets such that they have 3-point couplings with right-handed neutrinos at tree-level 
and their VEVs induce neutrino mass terms.
Majorana right-handed neutrino mass terms may be generated 
by higher dimensional operators\footnote{See for higher dimensional operators 
in magnetized brane models \cite{Abe:2009dr}.} and/or non-perturbative effects.
Indeed, non-perturbative computations to induce Majorana neutrino mass terms 
were studied in magnetized D-brane models \cite{Blumenhagen:2006xt,Hamada:2014hpa}.
Thus, it is quite interesting to apply such discussions for $SO(32)$ heterotic string theory.
We would study elsewhere.

\subsection*{Acknowledgement}
H.~A. was supported in
part by the Grant-in-Aid for Scientific Research No. 25800158 from the
Ministry of Education,
Culture, Sports, Science and Technology (MEXT) in Japan. T.~K. was
supported in part by
the Grant-in-Aid for Scientific Research No. 25400252 from the MEXT in
Japan.
H.~O. was supported in part by a Grant-in-Aid for JSPS Fellows 
No. 26-7296.

\end{document}